# Theoretical Study of Iridium-based PDT Photosensitizers for Improving Two-Photon Absorption, Triplet Lifetime and Lipophilicity through Ligand Tuning

*Aynur Matyusup[a], Jia-ying Zhao[b], Yu-dan Zhang[b], Qi Zhao[a], Ai-min Ren*[b], Jing-fu Guo*[a]*

[a]School of Physics, Northeast Normal University, Changchun 130024, P.R.China

[b]Institute of Theoretical Chemistry, College of Chemistry, Jilin University, Liutiao Road #2, Changchun 130061, P.R.China

**Abstract**

Iridium-based photosensitizers have attracted significant attention in photodynamic therapy (PDT) due to their exceptional photophysical properties and chemical stability, as well as tunable phosphorescence emission spectrum and high triplet state production yields. Photosensitizers with large two-photon absorption (TPA) and mitochondrial targeting capabilities are particularly promising for clinical PDT, as they enable deeper tissue penetration and reduced damage to normal cells. In this study, we theoretically studied photophysical, photodynamic properties and photosensitization reaction mechanism of a series of iridium-based photosensitizers with modified

C^N and N^N ligands (a2-a6, b1/b1-r and b2/b2-r) by TDDFT/DFT methods. The photophysical properties, including one- and two-photon absorption spectra, frontier molecular orbitals, and singlet and triplet excitation energies, were calculated. Additionally, rate constants for intersystem crossing, fluorescence, and phosphorescence, along with water solubility and lipophilicity metrics (logP), were determined to assess both efficacy and biocompatibility. The results elucidate the modulation roles of the chelated ligands and ancillary ligands in TP-PDT efficiency, indicating that the asymmetric iso-fused-benzene ring modification to the N^N ligand is a robust design strategy for comprehensively enhancing photosensitization performance. Complexes **a2**, **b2** and **b1-r** show greater promise as candidates for two-photon PDT photosensitizers, owing to their large TPA cross-sections, extended triplet state lifetimes, and balanced water solubility and lipophilicity. Notably, the **b1-r** complex can undergo both Type I and Type II PDT photosensitization mechanisms, which will help address the issue of drug resistance arising from the hypoxic environment in deep-seated tumors.

## 1.Introduction

Photodynamic therapy (PDT) has emerged as a minimally invasive and highly selective cancer treatment modality [1–3]. The core material for PDT is photosensitizer (PS), ideally a non-toxic molecule that has a higher affinity for cancer cells than for healthy cells when excited by light.

When irradiated with light, the PS molecule reaches a triplet excited state via intersystem crossing [4]. Photosensitizers in the triplet state generate cytotoxic reactive oxygen species (ROSs) through type I or type II mechanisms. In the former, activated triplet PS transfers an electron/H atom to an oxygen or organism substrate to form various ROSs such as superoxide anion radicals and hydroxyl radicals. During type II reaction, Photosensitizers transfer energy directly to ground-state oxygen (triplet state), creating highly energetic reactive singlet oxygen, $^1O_2$ [5–7].These ROSs induce oxidative damage to cellular tissues, leading to apoptosis or necrosis of cancer cells.

Overall, an effective photosensitizer for PDT should meet several fundamental requirements. First, its absorption wavelength is preferably within 600–900 nm, where light exhibits deeper tissue penetration and reduced damage to healthy tissues; (ii) an efficient intersystem crossing from the $S_1$ to the $T_1$ state is essential for sufficient production of $^1O_2$ or other ROS; (iii) the $T_1$ state should have a long lifetime for PS molecules to full react with ground state oxygens or self-ionization [8–11]. In addition to these basic requirements, an excellent photosensitizer should also be chemically stable, exhibit low dark toxicity, have minimal side effects, and be highly photostable to avoid photobleaching. (iv) it should be strongly phototoxic to effectively target cancer cells and biocompatible to ensure safety in biological systems[12].

To date, many different types of photosensitizers have been developed to improve treatment efficiency and versatility of photodynamic therapy (PDT). These include organic molecule-based photosensitizers, such as porphyrins and their derivatives, nanoscale materials like metal-organic frameworks (MOFs), and transition metal complexes-based PSs, including ruthenium, platinum, and iridium complexes [13–15]. Among these, iridium (III) complexes-based photosensitizers[16] have garnered significant attention due to their exceptional photophysical properties and chemical stability, including the following: (1) modular synthesis that allows easy preparation and modification; (2) the strong spin–orbit coupling of the Ir(III) center, leading to enhanced intersystem crossing and triplet-state accessibility; (3) first excited triplet states with extended lifetimes that can be deactivated by $^3O_2$ to generate $^1O_2$ or alternative ROSs; (4) a well-balanced hydrophilic/lipophilic character, combined with efficient cellular internalization and controllable intracellular localization, which can be tuned through ligand engineering; (5) phosphorescent emission with pronounced Stokes shifts that minimize concentration-induced self-quenching, thereby supporting bioimaging-assisted therapeutic protocols and theranostic uses and (6) robust photostability suitable for repeated excitation and multiple treatment cycles [17–23]. The photophysical properties of Ir (III) complexes can be precisely regulated through ligand modification. This tunability enables optimization of their absorption and emission wavelengths, allowing these complexes to be applied to various cancer treatments at different tissue depths. Such design flexibility facilitates the development of photosensitizers tailored to specific medical needs and therapeutic applications. In addition, iridium(III) complexes can be rationally engineered to target specific cellular structures or tissues [24,25].

To obtain Ir (III) complexes with long absorption wavelength (fall in the near infrared region), researchers utilized π-conjugation extension to optimize the photophysical and photodynamic properties of Ir (III) complexes, highlighting the feasibility of this approach as a pivotal strategy for the development of high-performance photosensitizers with enhanced PDT efficacy and diagnostic potential. In 2017, Wang and McFarland reported a series of near-infrared (NIR)-emitting heterolytic cationic iridium (III) complexes featuring 2,3-diphenylbenzo[g]quinoxaline ligands. These complexes demonstrated strong NIR phosphorescence, efficient reactive oxygen species (ROS) generation, and selective mitochondrial targeting, making them highly effective theragnostic agents for photodynamic therapy (PDT) [26]. Similarly, in 2022, Sanz-Villafruela and Massaguer designed Ir (III) complexes with extended π-conjugation on N^N ligands, resulting in red-shifted absorption and significantly enhanced singlet oxygen yields [18]. To address the issues such as limited tissue penetration depth and spatial resolution in photodynamic therapy (PDT), researchers have developed iridium (III) complexes with two-photon absorption (TPA). For instance, in 2023, Wang et al. developed two iridium (III) complexes for two-photon photodynamic immunotherapy (TP-PDI) against melanoma. Among them, Ir-pbt-Bpa showed a two-photon absorption maximum at 750 nm (75 GM), high photostability, efficient ROS generation, and induced cell death via ferroptosis and immunogenic cell death (ICD) [27]. In 2023, Li et al. developed cyclometalated iridium (III) complexes as two-photon photosensitizers for photodynamic therapy (PDT). These complexes showed two-photon absorption cross-sections of 102–391 GM at 808 nm. Among them, YQ2, containing piq as the **C^N** ligands, was the most potent infrared two-photon photosensitizer, with the highest singlet oxygen generation efficiency and strong mitochondrial localization[28]. To enhance the therapeutic efficiency and the safety of the PDT process, a variety of organelle-targeted Iridium (III) complexes have been developed. In **2016**, Qiu et al. showed that the organelle specificity of biscylometalated Ir(III) complexes can be tuned by ligand lipophilicity: the more lipophilic Ir1 and Ir2 (log P = 1.09 and 0.70) selectively accumulated in mitochondria, whereas the more hydrophilic Ir3–Ir5 (log P= −0.16 to −0.51) localized in lysosomes, with Pearson's colocalization coefficients of 75–87%[29]. In **2022**, Kuang et al. reported a mitochondria-targeted Ir(III) endoperoxide complex (2-O-IrAn) that displayed excellent mitochondrial selectivity (PCC = 0.96) and induced marked mitochondrial dysfunction, including membrane potential collapse and ROS generation upon light activation[30].

Although substantial progress has been made, there are currently very few types of Ir (III) complexes used as photosensitizers, and the actual clinical use of Ir (III) complex photosensitizers for PDT is very limited. The main problem is that the mechanism of action of these photosensitizers has not been fully elucidated, resulting in issues such as passivation of photosensitive reactions under hypoxic conditions, poor adaptability to the tumor microenvironment, and limited targeting efficiency that have not been fully resolved. Especially, the types of two-photon photosensitizers based on Ir (III) complexes are even rarer, and the absorption wavelength and intensity of two-photon sensitizers suitable for PDT still need improvement. It is urgent to develop effective design strategies for photosensitizers based on trivalent iridium complexes. In this context, McKenzie and co-workers reported two low-molecular-weight iridium photosensitizers capable of mitochondrial and lysosomal targeting. Among these, the complex $[Ir(N^\wedge C)_2(N^\wedge N)]^+$ demonstrated promising photophysical properties towards ideal photosensitizer, including a triplet excited-state lifetime of 1.9 μs, a two-photon absorption (TPA) cross-section of 112 GM, a high singlet oxygen quantum yield (0.42), mitochondria targeting ability with a Pearson's correlation coefficient of r = 0.547, and low dark toxicity [31]. Although this complex structure is effective in inducing cancer cell death under two-photon excitation (TPE) and exhibits great potential in two-photon photodynamic therapy (TP-PDT), there remains still considerable room for improvement in terms of its maximum TPA cross-section, targeting efficacy and triplet excited-state lifetime to achieve the desired photosensitization effect. Based on literature [32], the metal-iridium complexes with increased π-system of C^N ligand, such as 1-phenylquino line (**piq**), 2-phenylquinoline (**pq**), difluorophenylpyridine (**dfppy**), 2-(2-thienyl)pyridine (**thpy**), 2-phenylbenzothiazole (**pbt**), show more enhanced two-photon absorption, biocompatibility and lipophilicity than the complex with 2-phenyl pyridine (**ppy**) ligand. These modifications enhance mitochondrial targeting and, consequently, boost the therapeutic efficacy in photodynamic therapy (PDT). Benzimidazole, a fused heterocyclic system developed for therapeutic use since 1962, plays a significant role in antitumor, antimicrobial, anti-HIV, and antioxidant activities. Studies show that combining benzimidazole with larger conjugated aromatic rings in Ir(III) complexes enhances their TPA cross-sections, quantum yields for singlet oxygen ($^1O_2$) production and lipophilicity [23,25,33–36]. Espino et al. reported that increasing the number of aromatic rings in the N^N ligands of metal Ir(III) complexes leads to a reduction in the HOMO – LUMO band gap, a red-shift in the absorbance bands, prolonged excited-state lifetimes, and enhanced mitochondrial-targeting ability [37].

Based on above these researches, we take **[Ir(N^C)$_2$ (N^N)]$^+$** (denoted as **a1** here) as essential reference to further optimize the structure and performance of the Ir(III)-based complex as photosensitizers(PS) through theoretically simulation in this study. This aims to provide effective molecular design strategies for excellent candidates and appropriate theoretical simulation methods for experimental synthesis of clinically practical photosensitizers. For this purpose, we designed a series of Iridium (III) complexes (shown in **Figure 1)** through systematic ligand modifications. Specifically, ***a***-class molecules were derived by altering the chelating ligands directly coordinated to the metal center while keeping the **N^N** ligand unchanged; In contrast, ***b***-class molecules were designed by modifying the ancillary ligands with the primary ligand **C^N** part remaining unchanged. Using first principles calculations, we theoretically assessed critical parameters for PS molecules including one- and two-photon absorption properties, HOMO–LUMO energy levels, excited-state lifetimes, intersystem crossing (ISC) rates, solubility, lipophilicity and ROS generation mechanisms. Our ultimate goal is to elucidate the roles of the chelated ligands and ancillary ligands in modulating TP-PDT efficiency and to establish rational design strategies for novel Ir (III) complexes that exhibit enhanced photodynamic efficacy, prolonged triplet-state lifetimes, and precise subcellular targeting capabilities. This will provide a robust theoretical framework for future experimental exploration.

**Figure1.** The structures of experimental complex a1 and the new designed complexes.

## 2. Computational details

The calculations were performed via Gaussian 16 program [38] at Density Functional Theory (DFT) and Time-Dependent Density Functional Theory (TD-DFT). The absorption spectra of complex **a1** were calculated by several functionals, including B3LYP (HF = 20%), CAM-B3LYP [39], M06-2X (HF = 54%) [40], and PBE0 (HF = 25%) [41], and results were compared with the experimental value in order to screen the functionals for better describing the systems. The 6-31G(d) basis set was applied to nonmetallic atoms, while the SDD pseudopotential basis set was used for the Ir atom. Solvation effect was simulated using the polarizable continuum model (PCM) [42]in the dichloromethane ($CH_2Cl_2$) solvent. The results are presented in **Table S1**. Benchmarking results indicated that the PBE0/6-31G(d)/SDD method provided a suitable computational level for the studied systems. Consequently, the structures and properties of ground state and excited states for all complexes were calculated at this level of theory. Furthermore, the TPA wavelengths, TPA cross-section were obtained by the DALTON program via quadratic response theory [43,44]. Here, B3LYP functional was combined with the 6-31+G (d) basis set for C, H, O and N atoms and with the SDD pseudopotential basis set for Ir atom. The radiative decay rate and internal conversion rate were calculated by the MOMAP program. The SOC constants were calculated by the PySOC software [45,46]. Besides, vertical electron affinity (VEA) and vertical ionization potential (VIP) are computed at PBE0/6-31+G(d)/SDD level of theory.

## 3. Results and discussion

### 3.1. Geometrical structures of complexes

Geometric structure is essential for understanding the electronic structure and photophysical properties of the complexes. To investigate the structural changes between the $S_0$ and $T_1$ states, we calculated the geometrical parameters, including the bond lengths, bond angles and dihedral angles, as shown in **Table 1** and **Table S2**. Clearly, for the $S_0$ state, the bond lengths of **Ir−N1** and **Ir−N3**(refer to **Fig.1**) in complexes **a2** and **a3** are significantly longer than those in **a1.** This indicates that substituting 2-phenyl pyridine (**ppy**) with 1-phenylquinoline(**piq**), and 2-phenylquinoline(**pq**) in the **C^N** ligand weakens the **Ir−C^N** and **Ir−N^N** interactions. The reduction effect of 2-**pq** ligand modification in **a3** was more significant. Conversely, the bond lengths of **Ir−N1** and **Ir−N3** in complex **a4** are significantly shorter than those in **a1**, indicating that incorporation of difluorophenylpyridine (**dfppy**) into the **C^N ligand** strengthens both **Ir−C^N** and **Ir−N^N** interactions. Comparatively, in complexes **a5** and **a6(Fig. S2)**, the bond length of **Ir−N1** is shortened while that of **Ir−N3** is longer than in **a1**, indicating that replacing **ppy** with 2-(2-thienyl) pyridine (**thpy**) and 2-phenylbenzothiazole (**pbt**) enhances the **Ir−C^N** interaction while simultaneously weakening the **Ir−N^N** interaction. Furthermore, in complexes **b1**, **b2**, **b1-r**, and **b2-r**, the **Ir–N1** bond length is slightly shorter in **b1** and **b2**, whereas it is slightly longer in **b1-r** and **b2-r** compared to **a1**, while the **Ir–N3** bond length remains nearly unchanged across all b- series complexes. This indicates that such ligand modifications in b-series complexes primarily affect the **Ir−N^N interaction** with minimal impact on the **Ir−C^N interaction**. With respect to the bond angles, **C6–Ir–C5** and **N1–Ir–N2** in these complexes exhibited no significant variations compared to complex **a1**.

**Table 1.** Selected Geometrical Parameters of the Complexes (a1~ a**4**) in the ground state ($S_0$) and lowest triplet excited state ($T_1$).

| | a1 | | a2 | | a3 | | a4 | |
|---|---|---|---|---|---|---|---|---|
| | $S_0$ | $T_1$ | $S_0$ | $T_1$ | $S_0$ | $T_1$ | $S_0$ | $T_1$ |
| Bond lengths [Å] | | | | | | | | |
| Ir–N1 | 2.207 | 2.169 | 2.208 | 2.208 | 2.248 | 2.264 | 2.199 | 2.167 |
| Ir–N2 | 2.207 | 2.169 | 2.212 | 2.218 | 2.248 | 2.242 | 2.199 | 2.167 |
| Ir–N3 | 2.061 | 2.062 | 2.062 | 2.064 | 2.106 | 2.128 | 2.060 | 2.061 |
| Ir-N4 | 2.061 | 2.062 | 2.061 | 2.034 | 2.106 | 2.023 | 2.060 | 2.061 |
| Bond angle [deg] | | | | | | | | |
| C6–Ir–C5 | 89.4 | 88.4 | 89.4 | 89.9 | 90.0 | 89.5 | 88.8 | 88.0 |
| N1–Ir–N2 | 74.7 | 76.9 | 74.6 | 74.7 | 73.9 | 73.9 | 75.0 | 76.9 |
| Dihedral angle [deg] | | | | | | | | |
| N1–N2–Ir–N3 | -96.7 | -96.5 | -97.5 | -97.7 | -105.8 | -104.8 | -96.8 | -96.5 |

As for dihedral angles, relative to complex **a1**, the dihedral angle **N1–N2–M–N3** decreased by **9°** and **11.4°** in complexes **a5** and **a6**, respectively, which may be attributed to the Jahn–Teller effect. While it increased by **9.1°** in complex **a3**. In contrast, the **N1–N2–M–N3** dihedral angles in complexes **a2**, **a4**, **b1**, **b2**, **b1-r** and **b2-r** showed no significant changes.

In the $T_1$ state, the trends in changes of bond lengths, bond angles, and dihedral angles are essentially similar to those in the $S_0$ state. Overall, all complexes showed minimal structural changes between the $S_0$ and $T_1$ states, especially **a2, a5, a6, b1,** and **b2** exhibit minimal structural differences between the $S_0$ and $T_1$ states, suggesting limited structural relaxation upon excitation. Such small geometry variations are commonly associated with relatively higher molecular rigidity, which may suppress non-radiative decay pathways and thereby contribute to prolonged triplet lifetime [11,47,48].

In summary, geometric analysis reveals two distinct ligand-engineering paradigms. For C^N ligand modifications, the bulky extension in **a2**-**a3** elongates metal–ligand bonds, potentially weakening MLCT character, whereas the strong electron-withdrawing dfppy group in **a4** shortens them, synergistically enhancing both Ir–C^N and Ir–N^N interactions and thus favoring MLCT. In contrast, N^N ligand π-extensions maintain remarkably rigid coordination bonds in both $S_0$ and $T_1$ states. However, they subtly modulate the Ir–C^N interaction: the fused-ring extensions (**b1–**

**b2**) slightly enhance it in the excited state, while the iso-fused-ring extensions (**b1-r, b2-r**) induce a slight decrease. This indicates that N^N modifications primarily fine-tune photophysical properties through remote electronic effects without compromising structural stability, enabling precise performance optimization.

### 3.2. One-Photon Absorption Spectra

To investigate the impact of ligand modifications on the photophysical properties, the electronic absorption spectra of the newly designed complexes were calculated using the DFT//PBE0/6-31G (d, p)/SDD level of theory, the solvation effect in dichloromethane solvent also was considered by polarizable continuum model (PCM) in calculation. The electronic absorption spectra are shown in **Figure 2** and key one photon absorption (OPA) parameters, such as absorption maxima and longest absorption peaks and oscillator strengths, are listed in **Table S3**, which are essential for optimizing PDT efficacy.

In order to gain further insight into the electronic characteristics underlying the absorption behavior, we also calculated the frontier molecular orbitals (FMOs) of the complexes. The maps of the HOMOs (highest occupied molecular orbitals) and LUMOs (lowest unoccupied molecular orbitals), along with their energy gaps, are presented in **Figure 3**.

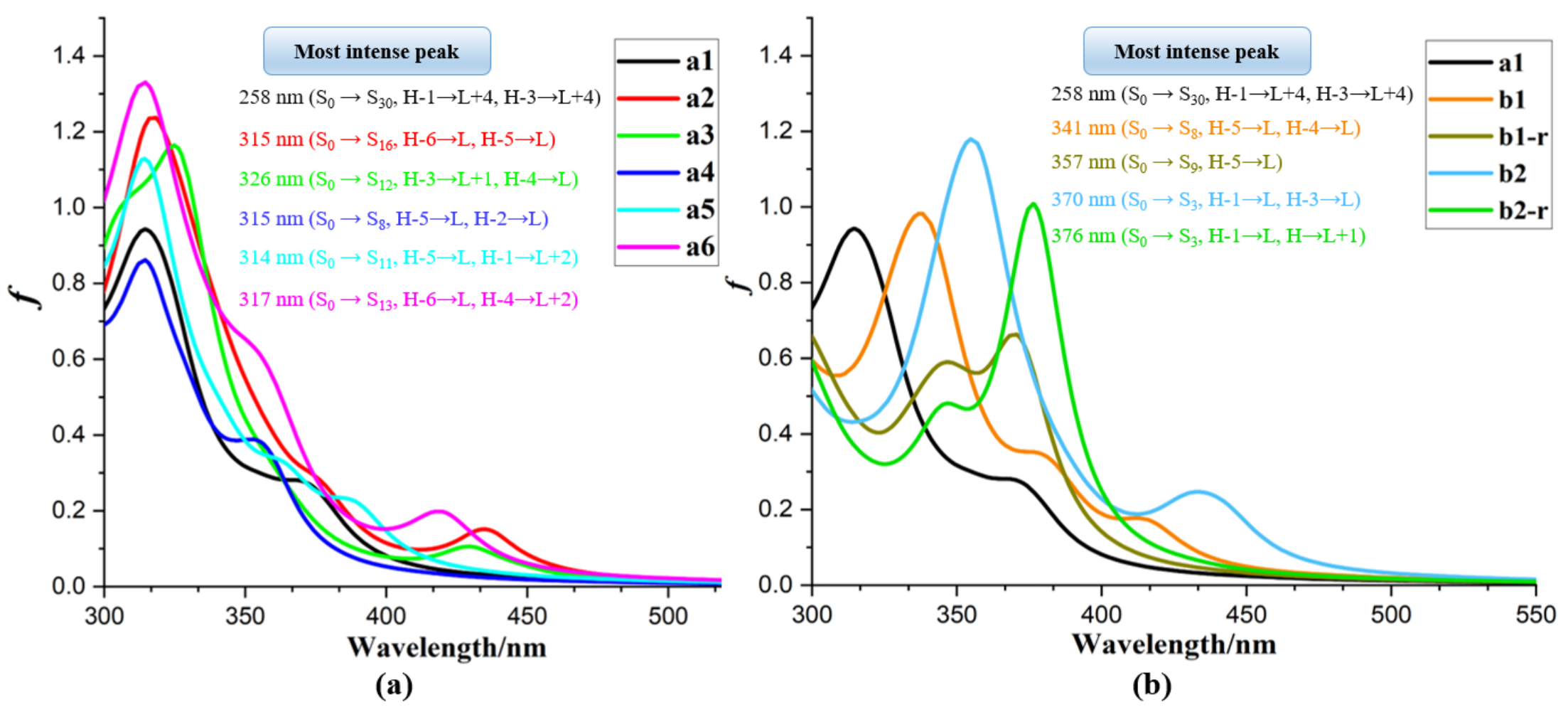

**Figure 2.** Calculated absorption spectra of **a1** to **a6** and **b1** to **b2-r** by using PBE0/6-31G (d, p)/SDD in dichloromethane. (a) a-class molecules. (b) b-class molecules.

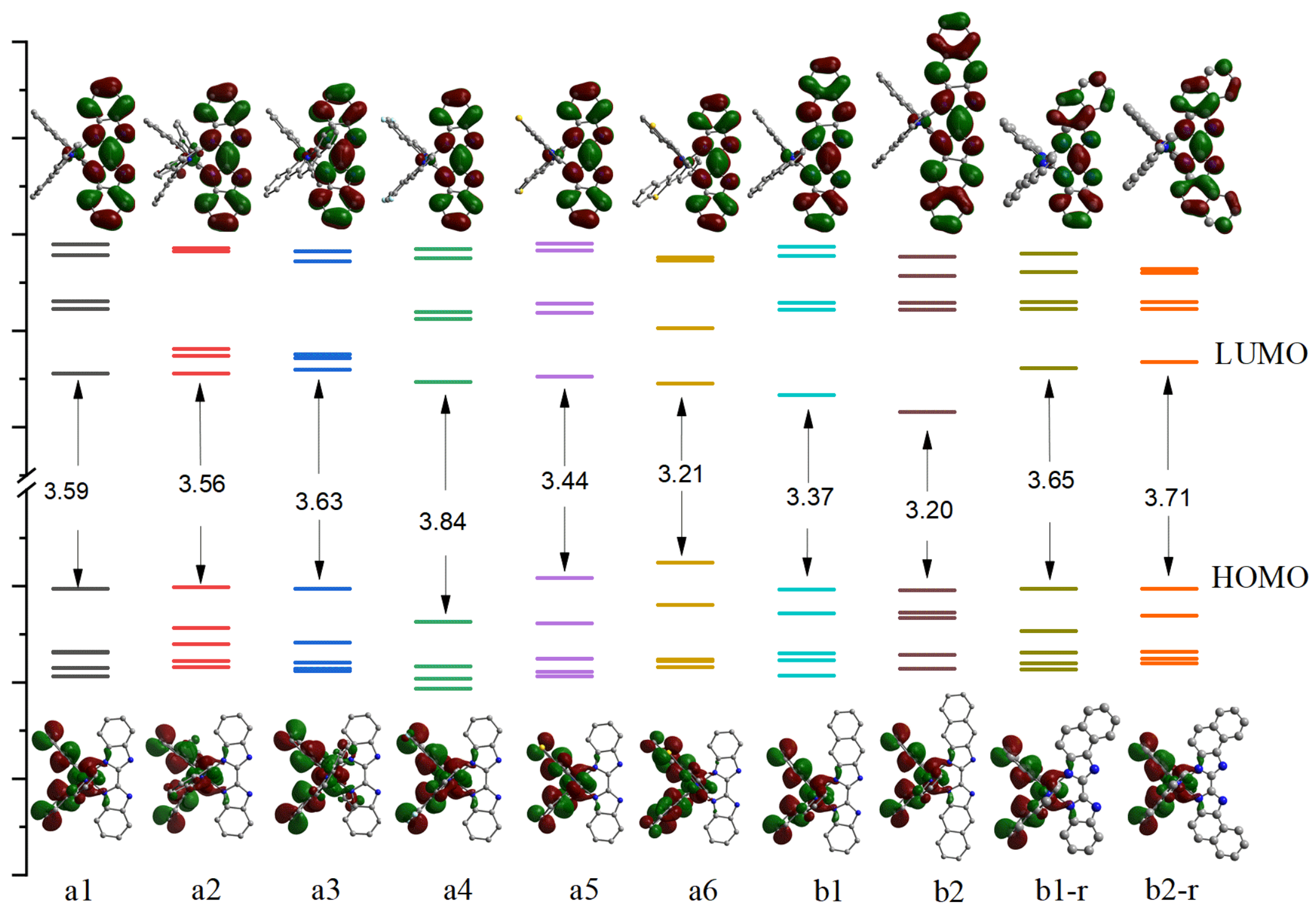

**Figure 3.** The Frontier molecular orbital maps, HOMO, LUMO and their energy levels, H-L gaps of the studied complexes.

As shown in **Figure 3** for **a**-series of molecules, the modification with the fused benzene ring to pyridine in the C^N ligand in compounds **a2** and **a3** results in the denser HOMOs' levels and LUMOs' levels compared to **a1**. In compound **a4**, the -F substitution at benzene ring in the C^N ligand markedly reduces the HOMO energy level and slightly reduces LUMO, thereby increasing the H-L band gap. In **a5** and **a6**, the introduction of the heterocyclic thiophene and benzothiophene evidently raises the HOMO energy level, thereby reducing the HOMO-LUMO energy gap by 0.2 – 0.3 eV. For **b**- series molecules, the introduction of fused benzene ring in the N^N ligand for **b1/b2** predominantly lowers the LUMO energy level, significantly decreasing the H-L band gap by 0.2 – 0.4 eV. Interestingly, the fused iso-benzene modification in either **b1-r/b2-r** or **a3** elevates the LUMO energy level, thereby increasing the H-L band gap while unchanged the pi character of HOMO and LUMO. The diversity resulting from electronic structure modifications will inevitably lead to the differences in electronic spectra for the designed molecules.

From calculated electronic spectra as shown in **Table S3** and **Figure 2**, for the experimental complex (**a1**), the red-edge absorption peak is located at 452 nm, which is predominantly contributed by the HOMO → LUMO transition (99%). The HOMO is predominantly localized on the metal center and the **C^N** ligand, while the LUMO is mainly distributed on the N^N ligand (see **Fig 3** and **Table S4**). Therefore, this transition is classified as having a mixed metal-to-ligand charge transfer (MLCT) and ligand-to-ligand charge transfer (LLCT) character. The maximum absorption peaks for the complexes with different ligands (**a2–a6**) are 454 nm, 446 nm ,413 nm, 474 nm and 507 nm, respectively. The absorption peaks are all predominantly contributed by the HOMO → LUMO transition, with the configuration contribution greater than 92% for all compounds. Compared to **a1**, the ligand modifications in **a2** and **a3** result in a slight change to the HOMO-LUMO gap, so their red-edge absorption wavelengths are also slightly red-shift and blue-shift. However, the modification of the fused benzene ring into pyridine in the C^N ligand has the HOMO-1, HOMO-2 levels up-shift and LUMO+1, LUMO+2 levels down-shift compared to **a1.** This renders HOMOs (and LUMOs) levels denser, optimizing orbital overlap and enhancing transition strength. Although the red-edge absorption wavelength does not shift significantly, the oscillator strength of **a2** and **a3** increases significantly compared to **a1**. Concurrently, the ligand modifications in **a2** and **a3** directly result in a redshift in the strongest absorption peaks to 315 nm and 326 nm (258 nm for **a1**), respectively. For **a4**, the absorption spectrum shows blue-shift (414 nm) attributed to meta-site substitution of the fluorine atoms in the C^N ligand. The electron-withdrawing effect of fluorine significantly lowers the HOMO level, while the LUMO level changes slightly, resulting in a broadened HOMO-LUMO gap. For **a5,** replacing the benzene ring with the thiophene in C^N ligand raises the HOMO energy level. This is because thiophene moiety in the **C^N** ligands enhance the electron-donating ability of the ligands, raising the HOMO level. Similarly, introducing the benzothiophene group extends the conjugation length in the **C^N** ligands for **a6,** increasing the HOMO level even more while decreasing the LUMO level, thereby reducing the HOMO-LUMO gap. This results in a red shift in the red-edge absorption wavelength and enhancement in the oscillator strength. For the complexes **(b1, b2)**, the maximum absorption peaks (485 nm and 517 nm) significantly red shift, with oscillator strengths enhanced by 0.0010. This is due to the decreased LUMO level resulting from the addition of a fused benzene ring to N^N ligand. Conversely, for the complexes **(b1-r, b2-r)**, the maximum absorption peaks (429 nm and 440 nm) significantly blue -shift, with oscillator strengths enhanced by 0.0013 and 0.015. This is ascribed to the elevated LUMO level resulting from the fused iso-benzene ring modification in N^N ligand. The strongest absorption peaks for **b-**series complexes occur between 341 nm and 376 nm, with oscillator strengths (***f***) ranging from 0.4589 to 0.8469. Compared to **a1** which has absorption peaks at 452 nm ($f$ = 0.0005) and 258 nm ($f$ = 0.3306), the **b**-class molecules exhibit significantly higher oscillator strengths, indicating stronger absorption transitions. These changes can be attributed to the introduction of an iso-fused benzene ring into N^N ligand, extending the π-conjugation in the N^N ligands and electron-donated ability of N^N ligands. This lowers the LUMO energy level and narrows the gaps among the LUMOs and HOMOs, thereby have the strongest electronic transition occurring between the less deep HOMOs and LUMO.

In summary, the modifications with the fused benzene ring, thiophene or benzothiophene or fluorinated in the primary (C^N) and auxiliary (N^N) ligands enhances electron-donated ability of N^N ligand, hence the electronic absorption of the **a-** and **b-**series complexes, strengthens the absorption intensity, and redshifts the strongest absorption peaks, thereby optimizing its optical properties. However, the one-photon absorption wavelengths of these complexes (417~517 nm) fall outside the photodynamic therapeutic window (600–900 nm), [49,50] which could be an issue as photosensitizers because short-wavelength light (400 – 500 nm), such as ultraviolet and blue-violet light, is easily absorbed by pigments (such as melanin) on the surface of the skin and haemoglobin in the blood. This makes it difficult to reach deep-seated diseased tissues. Nevertheless, these complexes, with one-photon absorption wavelength of around 450 nm, may be suitable candidates for the two-photon excitation light sources at 780 nm and 860 nm. Consequently, in the subsequent section, we will assess their two-photon absorption (TPA) properties.

### 3.3. Two-Photon Absorption (TPA) Properties.

Generally, two-photon photosensitizers that can be excited by long-wavelength light and exhibit relatively large two-photon absorption cross-sections are considered advantageous for photodynamic therapy, as they enable deeper tissue penetration and reduced photodamage to normal tissues. In practical applications, a cross-section on the order of 100 GM or higher is often regarded as desirable for effective two-photon PDT. two-photon PDT [51,52].

The two-photon absorption (TPA) properties of the studied molecules were therefore evaluated by using the DALTON program [43,44], based on the optimized $S_0$ state geometry.

#### 3.3.1. Validation of the TPA Computational Protocol.

To assess the reliability of the computational protocol for comparative analysis, the reference compound **a1** (with available experimental TPA data) together with two structurally related

iridium(III) complexes **IrM4** [35] and **IrC2** [53], both reported by the same authors, were selected as validation systems (for detailed information, refer to Section II in the Supporting Information).

The TPA cross-sections were computed at the B3LYP/6-31+G(d)/SDD level using the same protocol applied throughout this work. As summarized in **Table 2,** the calculated absolute cross-sections are consistently underestimated by approximately one order of magnitude relative to experimental values (e.g., 13.72 vs 112 GM for **a1**; 35.83 vs 368 GM for **IrM4**; 28.05 vs 302 GM for **IrC2**). The underestimation of TD-DFT-calculated two-photon absorption (2PA) cross-sections relative to experimental values is a commonly reported issue in theoretical studies [54,55]. This discrepancy mainly arises from several factors. First, TD-DFT may underestimate excited-state dipole moments and charge-transfer character, particularly for metal complexes involving MLCT or ligand-centered charge transfer, which directly affects the predicted $\sigma_2$ values. Second, the calculations describe only the electronic contribution to the 2PA process, whereas experimental measurements may include additional enhancements from vibronic coupling. Third, environmental effects such as solvent polarization and local-field effects in condensed phases may further increase the experimental cross-sections [56–58].

**Table 2.** The two-photon absorption wavelengths and cross-section of a1, IrM4 and IrC2 complexes by B3LYP/6-31+G(d)/SDD, and the experimental values are presented in parentheses.

| **Mol.** | **$\lambda_{tpa}$/nm** | **δ/GM** |
|---|---|---|
| **a1** | 751(760) | 13.72(112) |

| IrM4 | 841(810) | 35.83(368) |
|---|---|---|
| IrC2 | 835(810) | 28.05(302) |

Importantly, the relative ranking among the complexes (**IrM4** > **IrC2** > **a1**) is accurately reproduced by the calculations, consistent with experimental observations. This confirms that, despite the systematic underestimation of absolute magnitudes, the trends in relative comparison remain consistent, allowing for a reliable assessment of the comparative performance of the designed molecules.

**3.3.2. TPA Properties of the Designed Complexes.**

Based on the trend reliability of the computational protocol, we next analyze the TPA properties of the designed complexes.

The maximum TPA cross-section and corresponding absorption wavelength, as calculated by B3LYP and CAM-B3LYP, are listed in **Table 3** and **Table S6**. Both methods produced consistent trends in the variation of the TPA cross-sections and showed similar differences compared to experimental values. However, the absorption wavelengths calculated with B3LYP were found to be in better agreement with the experimental data. Therefore, we based our analysis on the results obtained from B3LYP and focused on TPA transitions above 700 nm.

**Table 3.** Maximum TPA Cross-Section (δ/GM), Corresponding TPA Wavelength ($\lambda_{tpa}$, nm), and transition Character of the Complexes Calculated at the B3LYP/6-31+G(d)/SDD Level.

| Mol. | Energy/eV | $\lambda_{tpa}$/nm | δ/GM | Transition character | | Assignment |
|---|---|---|---|---|---|---|
| **a1** | 3.30 | 751 | 13.72 | $S_0 \rightarrow S_5$ | H-2→L (76.15%) | LLCT |
| | | | | | H-3→L (18.88%) | LLCT/ILCT |
| **a2** | 3.34 | 742 | 36.20 | $S_0 \rightarrow S_9$ | H-2→L+2 (73.06%) | LLCT/ILCT |
| | | | | | H-1→L+1 (10.88%) | ILCT |
| **a3** | 3.42 | 725 | 10.70 | $S_0 \rightarrow S_9$ | H-2→L+1 (76.19%) | LLCT |
| | | | | | H-1→L+2 (4.96%) | ILCT |
| **a4** | 3.88 | 639 | 11.36 | $S_0 \rightarrow S_{11}$ | H-2→L+1 (47.57%) | ILCT |
| | | | | | H-1→L+2 (30.67%) | LLCT |
| **a5** | 3.43 | 723 | 10.77 | $S_0 \rightarrow S_6$ | H-4→L (93.44%) | LLCT/ILCT |
| **a6** | 3.42 | 725 | 8.10 | $S_0 \rightarrow S_8$ | H-3→L (68.11%) | LLCT |
| | | | | | H-5→L (19.02%) | LLCT/ILCT |
| **b1** | 3.21 | 772 | 8.19 | $S_0 \rightarrow S_6$ | H-3→L (56.32%) | LLCT/ILCT |
| | | | | | H-2→L (20.57%) | LLCT/ILCT |
| **b2** | 3.20 | 775 | 15.53 | $S_0 \rightarrow S_8$ | H→L+2(75.95%) | ILCT |
| | | | | | H-6→L (16.15%) | LLCT |
| **b1-r** | 3.44 | 721 | 56.61 | $S_0 \rightarrow S_6$ | H-2→L (43.95%) | LLCT |
| | | | | | H-3→L (29.59%) | LLCT/ILCT |
| **b2-r** | 3.40 | 729 | 34.89 | $S_0 \rightarrow S_6$ | H-2→L (81.70%) | LLCT/ILCT |
| | | | | | H-3→L (7.65%) | LLCT/ILCT |
| | 3.77 | 657 | 219.00 | $S_0 \rightarrow S_{13}$ | H-4→L+1 (51.15%) | LLCT/ILCT |
| | | | | | H-2→L+2 (22.78%) | LLCT/ILCT |

Complex **a1** was experimentally reported to have a two-photon absorption cross-section of 112 GM at 760 nm, efficiently inducing apoptosis in cancer cells under near-infrared light excitation, highlighting its potential as a two-photon photodynamic therapy photosensitizer [31]. The TPA wavelength calculated with B3LYP (751 nm) is in good agreement with the experimental value, although the maximum TPA cross-section is 13.72 GM, which underestimates the absolute TPA value. Further analysis reveals that the calculated TPA cross-sections are approximately underestimated by one order of magnitude at the B3LYP level (refer to **Table 2**).

As for the modified complexes, **a2**, **b1-r** and **b2-r** exhibit relatively larger calculated TPA cross-section of 36.20 GM, 56.61 GM, 34.89 GM, respectively, at the wavelength of 742 nm, 721 nm and 729 nm. These calculated values are higher than that of the parent compound **a1** (13.72 GM at 751 nm) within the same computational framework. The observed increase appears to be related to the fused-benzene-ring modifications introduced in **C^N** or **N^N** ligands, which densify the frontier orbitals and render the added fused-benzene-ring participate in two-photon transitions, thereby increasing the number of two-photon transition configurations.

Meanwhile, **b2** shows a TPA cross-section of 15.53 GM at 775 nm, this is not too small, it benefits from extended π conjugation and improved molecular rigidity provided by the modified N^N auxiliary ligand. In contrast, compounds **a3, a4, a5, a6,** and **b1** display moderate TPA cross-sections ranging from 8.10 to 11.36 GM, with absorption wavelengths generally blue-shifted compared to **a1**. Although the TPA cross-sections of **a3–a6** and **b1** are somewhat lower than that of **a1**, they all fall within the therapeutic window (600–900 nm). For reference, the calculated δ values are numerically higher than the reported experimental values of several clinically used photosensitizers, such as **Photofrin** (2.2 GM) [59] and **[Ru(bpy)$_3$]$^{2+}$** (4.3 GM) [60] ; however, direct quantitative comparison should be treated with caution due to methodological differences. Taken together, these findings suggest that, despite the relatively moderate δ values, these complexes exhibit computationally appreciable two-photon absorption responses within the present model.

To summarize the above analysis, it shows that the modification with fused-benzene/ iso-fused benzene ring to **C^N** ligands reduces the electron- donated ability of **C^N** ligands in two-photon transitions, and increases ILCT transition configuration from **C^N** ligand in **a2** and **a3**; while the modification with fused-benzene/iso-fused benzene ring to **N^N** ligands reduces the electron-accepted ability of N^N ligands in two-photon transitions, and make the two-photon transition strength change. It is noticed that iso-fused benzene ring modification to **N^N** ligand plays the role of electron-donor besides the reduction of electron-accepted property, rendering the ILCT contribution from **N^N** ligand to be significant in **b1-r** and **b2-r** and become the important configuration of two-photon transitions, thereby greatly increasing the two-photon absorption cross-section value. The modification of **C^N** ligands with thiophene and benzothiophene in **a5** and **a6** limitedly deducts the participation of **C^N** ligands in two-photon transitions. Meanwhile, it slightly increases the involvement of **N^N** ligand as the electron-donor in two-photon transition. Although this also increase the ILCT from **N^N** ligand, it is unhelpful to the increase in the two-photon absorption cross-section. And the modification with fluorine substitution at meta-site in **a4** substantially diminishes the electron-donating capacity of the **C^N** ligands upon the TPA transition. This results in a reversal of charge-transfer direction during the TP transition compared to the transition configuration of **a1,** the **C^N** ligands transform from an electron-donor to an acceptor, whilst the **N^N** ligand becomes an electron-donor, thereby reducing the two-photon absorption intensity of **a4**.

The enhancement of TPA responses induced by the asymmetric iso-fused benzene ring modification can be understood from an electronic-structure perspective. The fused aromatic framework extends the π-conjugation and increases molecular rigidity, which facilitates stronger electronic delocalization and stabilizes charge-transfer states. Meanwhile, the asymmetric modification modulates the frontier molecular orbital distribution and enhances ILCT/LLCT contributions in the dominant TPA transitions, thereby increasing the probability of two-photon transitions. These results suggest that introducing asymmetric π-extended fused aromatic units may represent a promising structural strategy for enhancing TPA responses in related photosensitizer systems. Taken together, these results suggest that **a2**, **b2**, **b1-r**, and **b2-r** exhibit relatively enhanced TPA responses within the validated computational framework.

### 3.4. Triplet State Population and Lifetime

Upon excitation by one or two-photon radiation, the photosensitizer molecules reach a singlet excited state. Whether excited photosensitizers can successfully populate the $T_1$ state is critical at this stage. This is because the triplet excited state ($T_1$) plays a crucial role in photodynamic therapy (PDT), as only photosensitizers that successfully populate the $T_1$ state can transfer energy or electrons to molecular oxygen, generating reactive oxygen species (ROS) as shown in Scheme 1. The longer the photosensitizer remains in the triplet state, the more concentration it has to react with oxygen, resulting in a higher production yield of ROS. Therefore, it is essential to understand the population and lifetime of the $T_1$ state for evaluating the PDT efficiency of photosensitizers.

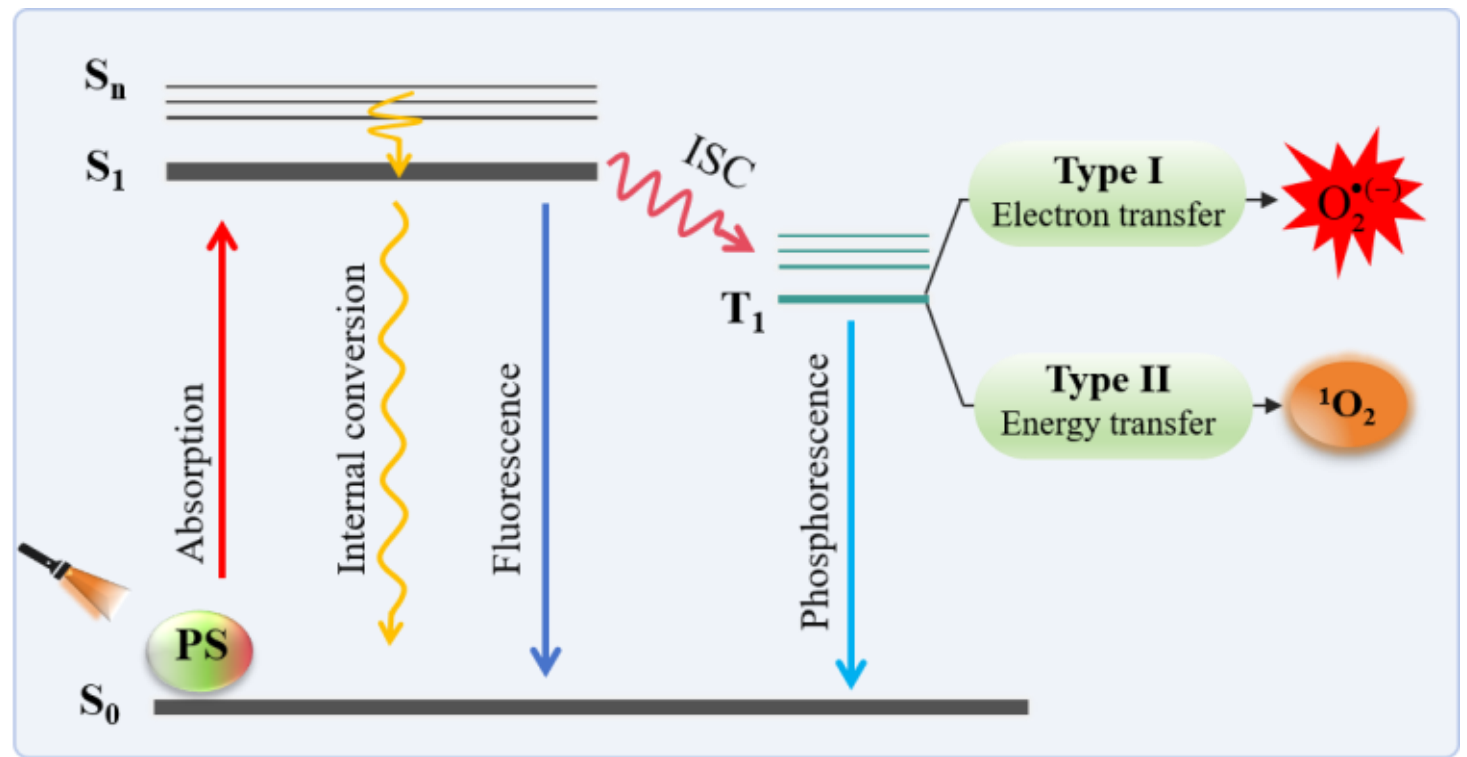

**Scheme 1.** The schematic diagram of the photophysical processes of photosensitizer

As is well known, larger SOC constants and smaller singlet-triplet energy gaps promote the ISC process. For it, we calculated the SOC constants and energy gaps between the $S_1$ state and $T_n$ states that are closest to the $S_1$ state for studied complexes. At the same time, we also calculated the SOC value of the clinically used photosensitizer Foscan at the same calculation level as a reference [63]. The calculated results are shown in **Figure4** and **Table S7**. All the studied molecules have only triplet excited state ($T_1$) with energy lower than that of the $S_1$ state except for the **b2**, which has two triplet excited states ($T_1$ and $T_2$) with the energies lower than that of the $S_1$ state.

As depicted in **Figure 4**, complexes **a1, a2, a3, a4, b1, b2, b1-r** and **b2-r** have the higher SOC constants and are 228.13 $cm^{-1}$, 204.77 $cm^{-1}$, 126.99 $cm^{-1}$, 269.0 $cm^{-1}$, 196.82 $cm^{-1}$, 174.77 $cm^{-1}$, 240.91 $cm^{-1}$, and 207.09 $cm^{-1}$, respectively, indicating that they are all more likely to exhibit stronger ISC. In contrast, complexes **a5** and **a6** exhibit relative lower SOC values of 73.13 $cm^{-1}$ and 12.25 $cm^{-1}$. These implies that the modification with the unsymmetrically fused/iso-fused-benzene ring to **N^N** ligand and meta-difluorine substitution to **C^N** ligand is more effective strategies to improve SOC value in comparison with **a1**. However, the SOC values of all the designed complexes are significantly higher at least 10 times than that of Foscan (calculated SOC constant is 1.03 $cm^{-1}$). Moreover, previous studies have shown that SOC values ranging from 0.2 to 5.0 $cm^{-1}$ are adequate to trigger ISC within nanoseconds [64]. Therefore, all the investigated complexes are expected to efficiently undergo intersystem crossing (ISC).

Upon light irradiation the PSs are excited to the $S_1$ or higher excited states Sn (highly Sn will relax to $S_1$ through a fast internal conversion (IC) process). There are two pathways for photosensitizers (PSs) in the $S_1$ state to deactivate. One is to return to the ground state ($S_0$) via a radiative transition process (fluorescence emission) or through nonradiative transition process (internal conversion). The other path is going to first excited triplet state ($T_1$) or higher triplet states ($T_n$) via the ISC process, followed by a rapid internal conversion to the lowest triplet state ($T_1$). Due to the competition of these two pathways, it is essential to quantitatively assess the kinetic transition behaviors of studied molecules to determine whether they can effectively reach the $T_1$ state or predominately populate $T_1$ state.

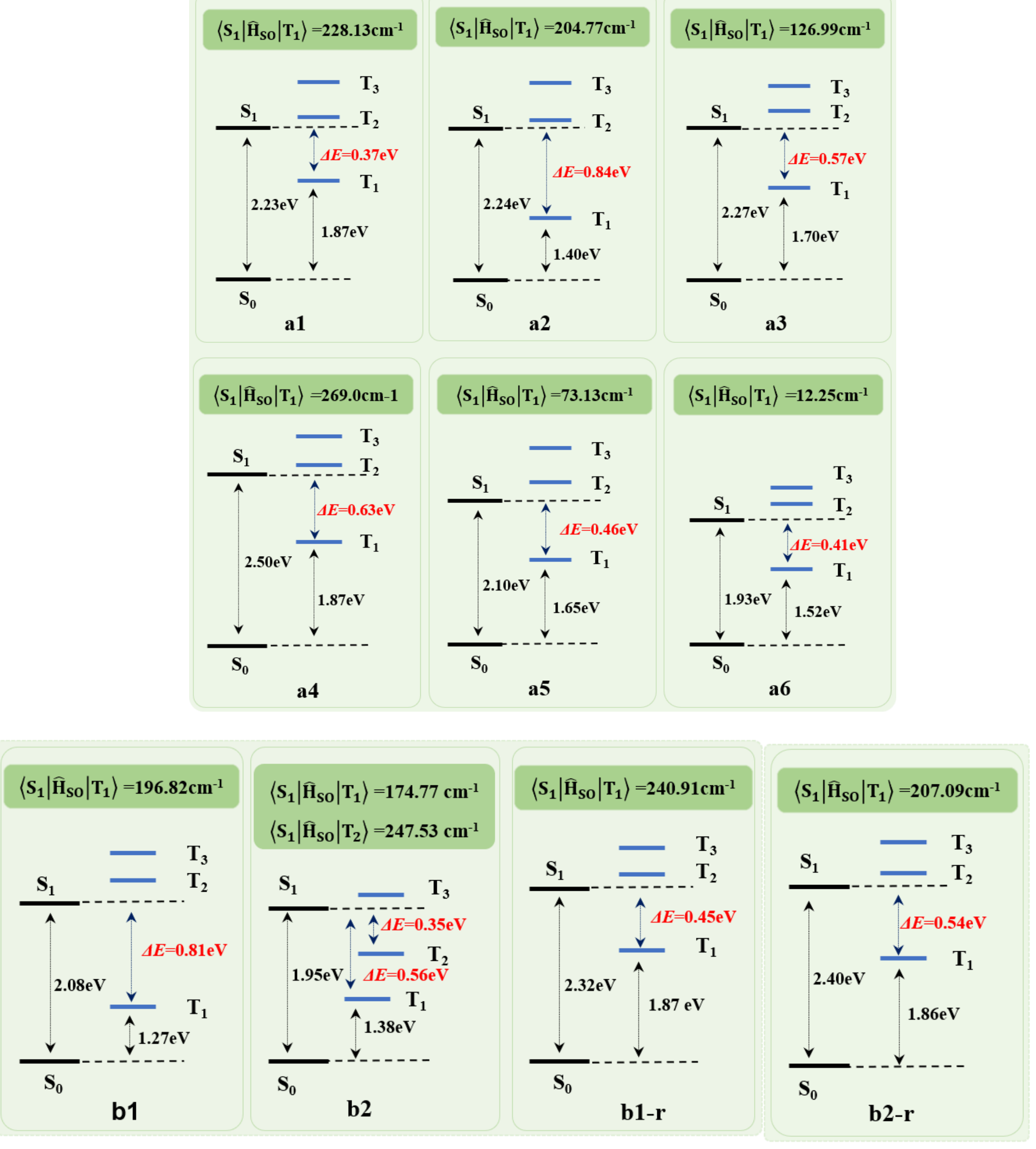

**Figure 4.** The energies of the first singlet state and triplet excited states and the orbital coupling constants of studied complexes.

For this purpose, the rate constants for internal conversion ($K_{ic}$), fluorescence ($K_f$), and intersystem crossing ($K_{isc}$) were calculated by the MOMAP program for all molecules based on the $S_1$ state geometries. The results are presented in **Figure 5**, and the specific values are listed in **Tables S7** and **S8**. As shown in **Table S7**, for the studied complexes, the largest $K_{isc}$ occurs in the $S_1 \rightarrow T_1$ transition besides the complex **b2**, which the largest $K_{isc}$ occurs in the $S_1 \rightarrow T_2$ transition. The maximum $K_{isc}$ values for all studied complexes ($3.69\times10^{10}$ ~$2.57\times10^{13}$ $s^{-1}$) are more competitive than both $K_f$ ($1.65\times10^{1}$ ~$4.44\times10^{5}$ $s^{-1}$) and $K_{ic}$ ($6.8\times10^{4}$ ~$2.31\times10^{11}$ $s^{-1}$), as shown in **Figure 5 and Table S7~S8**. Therefore, it is evident that the ISC process is the dominant transition pathway. On the basis of the above analyses, the main deactivation path for studied complexes is $S_0 \xrightarrow{\text{light abs.}} S_1 \xrightarrow{\text{ISC}} T_1$ except for **b2** is $S_0 \xrightarrow{\text{light abs.}} S_1 \xrightarrow{\text{ISC}} T_2 \xrightarrow{\text{IC}} T_1$. Although the $K_{isc}$values of the modified complexes are lower than those of the experimental molecule **a1**, all of their triplet quantum yields are all greater than 97% (see **Table S7**). As a result, these molecules exhibit relatively high triplet harvesting efficiency.

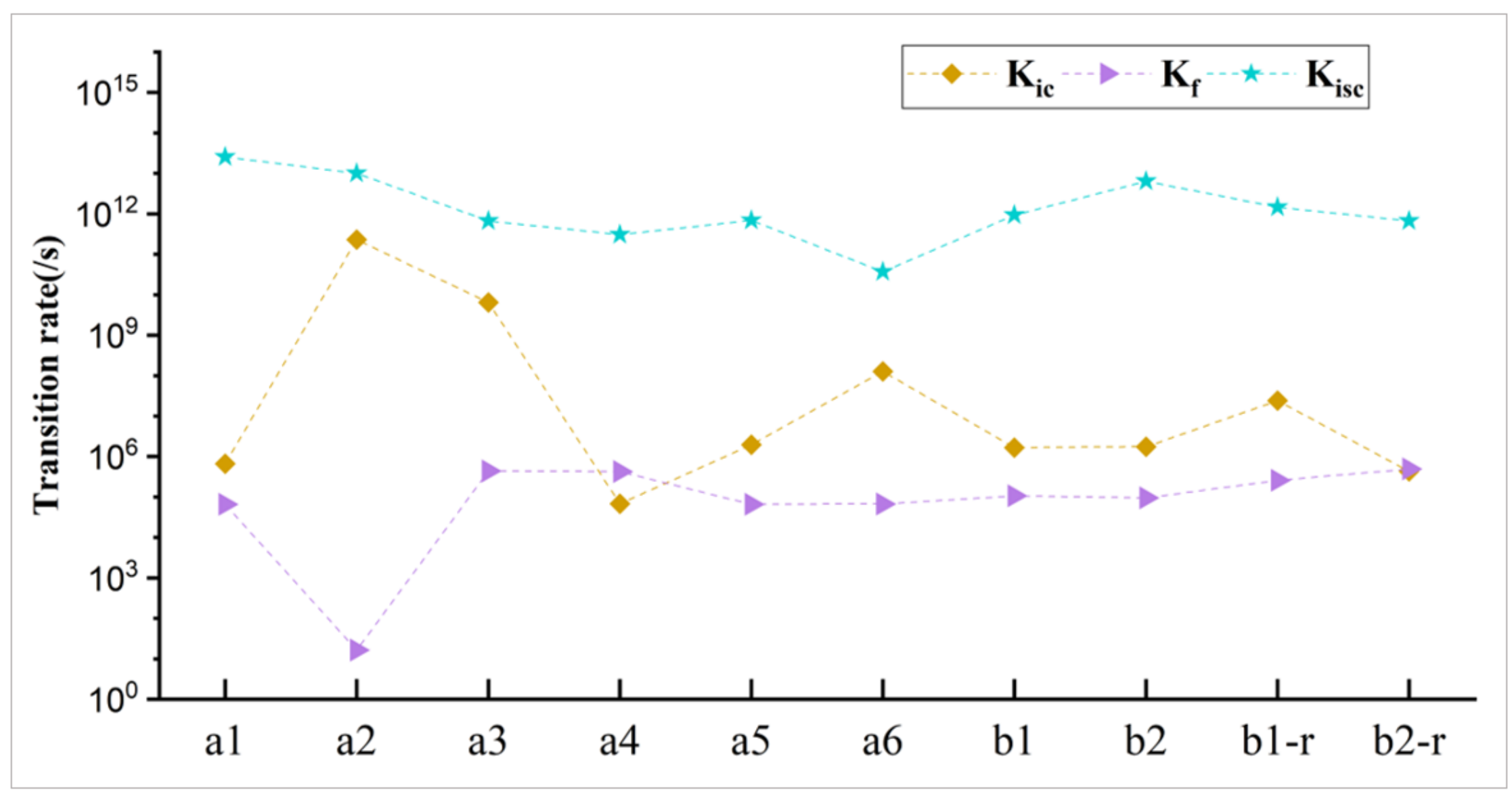

**Figure 5.** The rate constants for the three decay paths of the $S_1$ state: the fluorescence radiation decay rate $K_f$ ($s^{-1}$), the ISC rate $K_{isc}$ ($s^{-1}$), the internal conversion rate $K_{ic}$ ($s^{-1}$).

The natural lifetimes τ of triplet state of the studied complexes, which is defined as the reciprocal of $k_r$, were estimated in terms of calculating the radiative transition rate ($k_r$) of triplet state $T_1$. The results are shown in **Figure 6** and detailed in **Table S9**. For the experimental reference compound **a1**, the calculated triplet state lifetime is 2.19 μs, which is reasonably consistent with the experimental value (1.9 μs), supporting the reliability of the computational approach for relative comparison. The modified complexes exhibit a broad range of triplet lifetimes, from **0.87** μs to **85.96** μs. Notably, **a6** (85.96 μs), **a5** (71.64 μs), **b2** (31.06 μs), and **a2** (21.84 μs) display comparatively longer lifetimes, whereas **b1-r** (3.58 μs) and **b2-r** (0.87 μs) show shorter values. Except for **b2-r**, all designed complexes exhibit triplet lifetimes longer than that of the reference compound **a1.**

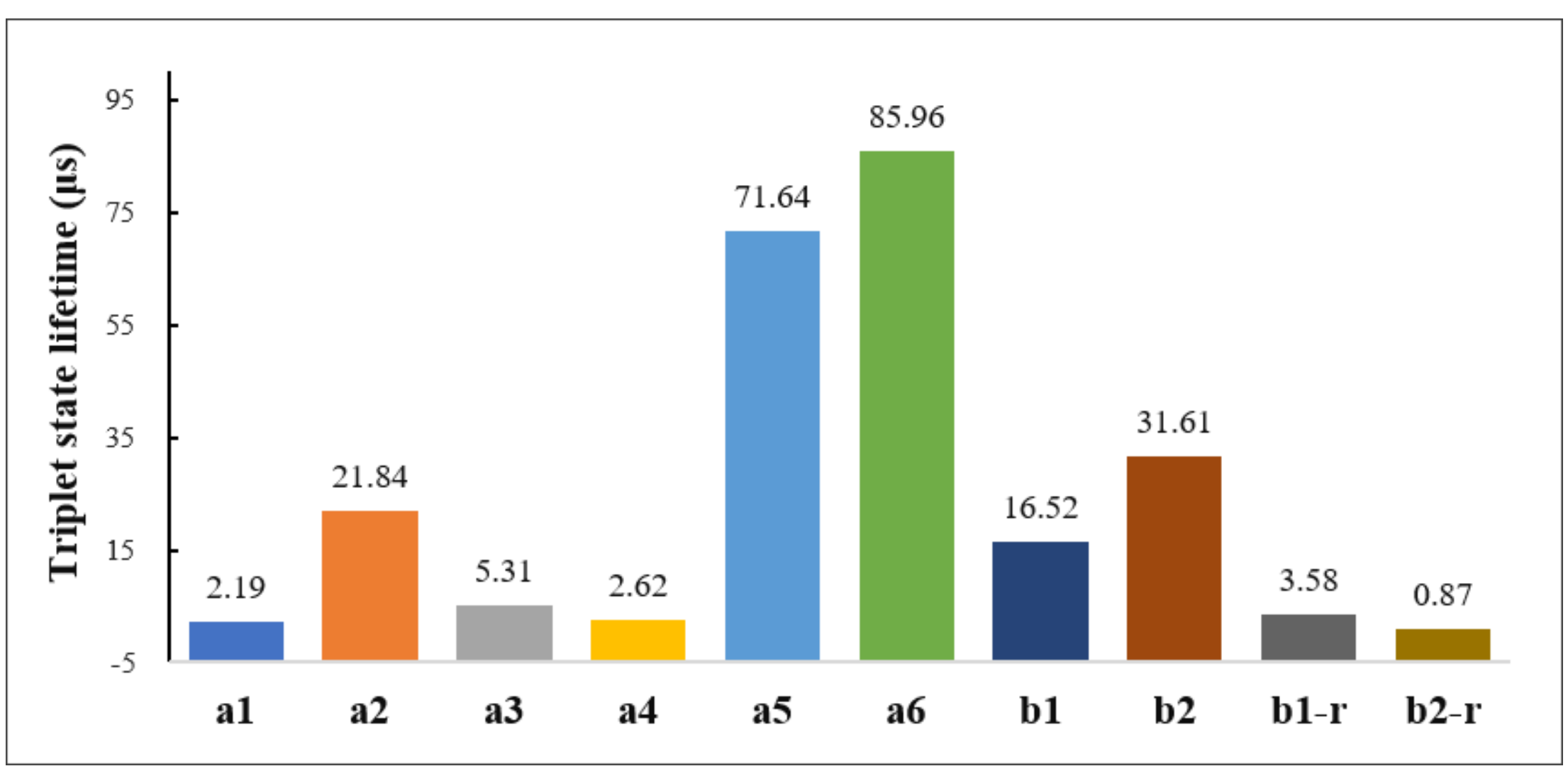

**Figure 6.** Calculated triplet state lifetime τ (μs) of the studied complexes.

To clarify how ligand modifications influence the triplet lifetimes, the SOC matrix elements, the vertical excitation energies of the $T_1 \rightarrow S_0$ transition, and the phosphorescence radiative rates between the $T_1$ and $S_0$ states were analyzed (**Table S9**). Incorporation of fused-benzene rings into the C^N ligand (**a2**, **a5**, and **a6**) is accompanied by markedly reduced SOC values relative to **a1**. In addition, **a2**, **a3**, **a5**, **a6**, **b1**, and **b2** exhibit lower vertical excitation energies for the $T_1 \rightarrow S_0$ transition compared to **a1**. According to **eq S3** (Supporting Information), the radiative decay rate depends on both the square of the SOC strength and the cube of the $T_1$ transition energy; thus, the combined reduction in SOC and $T_1$ transition energy is expected to decrease $K_r$, contributing to prolonged triplet lifetimes. Moreover, minimal structural differences between the $S_0$ and $T_1$ states for **a2**, **a5**, **a6**, **b1**, and **b2** suggest increased rigidity (see **Table S1**), which may suppress non-radiative decay pathways ($K_{nr}$). Collectively, these effects account for the lifetime variations induced by specific ligand modifications.

As noted in previous studies, the efficiency of a photosensitizer is closely related to its triplet state properties, including a high quantum yield **($\Phi_T$>0.4)** and long lifetimes **($\tau$>1 $\mu$s)** [65]. In general, longer triplet lifetimes can enhance the probability of interaction with molecular oxygen and thus favor reactive oxygen species generation. However, overall PDT performance is governed by multiple factors, including absorption efficiency, intersystem crossing characteristics and excited-state lifetime, as well as redox properties of photosensitizers involving type I/II mechanisms.

### 3.5. Type I and Type II Mechanisms

After confirming that the studied complexes can reach the triplet state with sufficient lifetime upon light radiation, the next step involves determining whether the triplet states transfer their energy or electrons to molecular oxygen, generating either high-cytotoxic singlet oxygen ($^1O_2$) via Type II mechanism or superoxygen anion radical ($O_2^{\bullet(-)}$) via Type I mechanism. This is crucial for the destruction of cancer cell.

To enable Type II mechanism, the triplet state energy must be above 0.98 eV to promote the $O_2$ from its ground state $^3\Sigma_g$ to the excited state $^1\Delta_g$ [63,66]. The triplet state energies of the studied complexes, illustrated in **Figure 7** and **Table S11**, all surpass this threshold, indicating their possibility of generating $^1O_2$ through energy transfer (Type II mechanism).

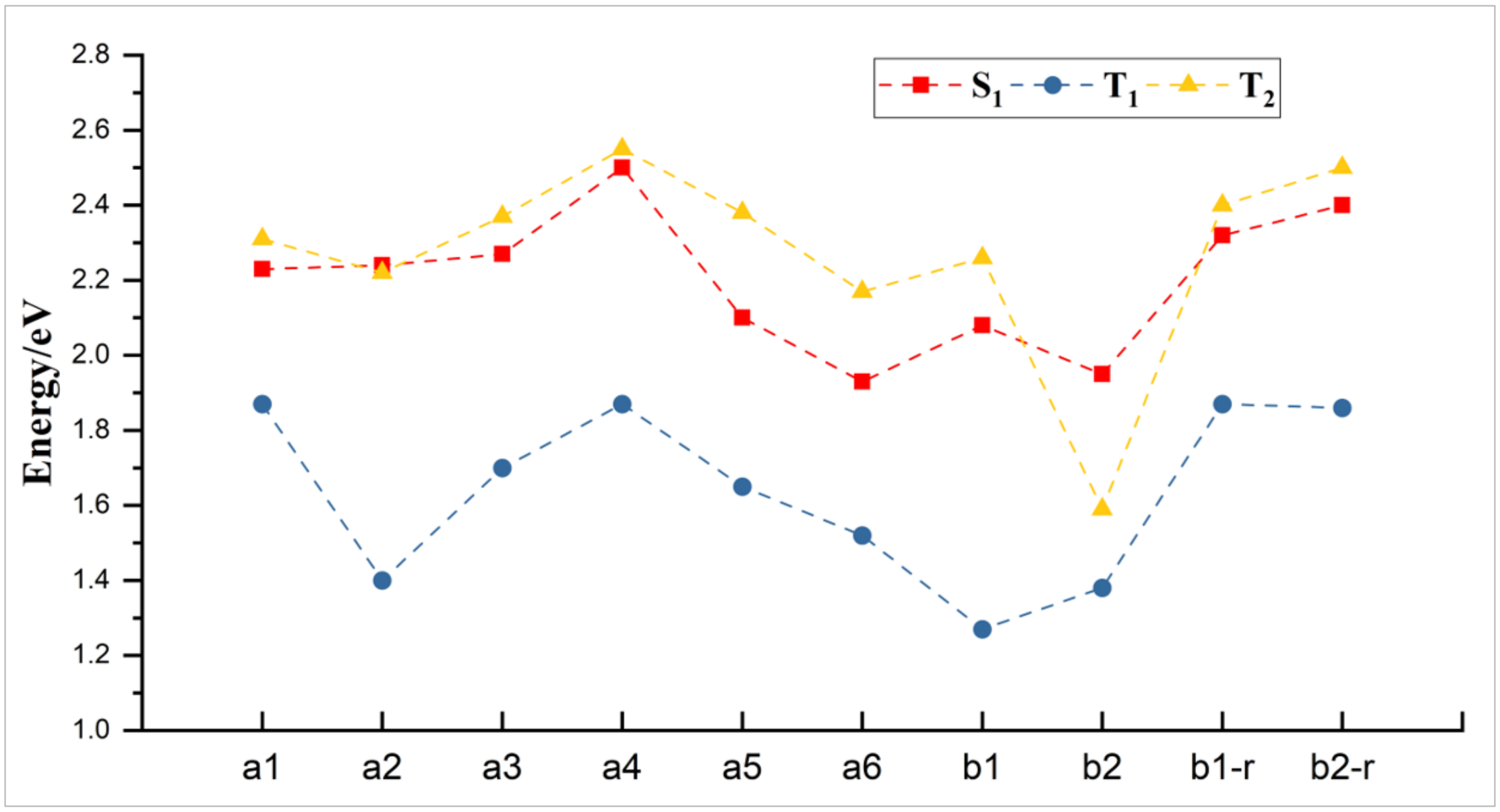

**Figure 7.** Calculated triplet excited state energies ($T_1$, $T_2$), singlet excited state energy ($S_1$) computed using PBE0/6-31G (d, p)/SDD in dichloromethane.

The populated excited triplet state PS could also interact with surrounding oxygen molecules through electron transfer, known as the Type I mechanism, generating cytotoxic reactive oxygen species (superoxygen anion radical, refer to the following reactions (1)-(4)). This pathway is particularly advantageous in hypoxic tumor microenvironments [5–7]. To assess whether the studied complexes can undergo Type I photoreactions, the Gibbs free energy of electron transfer reactions between the photosensitizer and oxygen must be considered. The vertical electron affinity (VEA) and ionization potentials (VIP) for each complex and molecular oxygen are calculated at PBE0/6-31G +(d)/SDD level in DMSO [50,67], and the results are collected in **Table 4**.

In Type I mechanism, the PS directly transfer electrons to molecular oxygen reaction and produce superoxide anion $O_2^{\bullet(-)}$ according to reaction (1):

$$PS + {}^{3}O_2 \rightarrow PS^{\bullet(+)} + O_2^{\bullet(-)} \qquad (1)$$

To ensure the reaction happens the sum of VIP (PS, $T_1$) and VEA($^3O_2$) must be smaller than zero. The computed data shows that none of the studied molecules meet the conditions required for reaction (1).

The other possible way to generate $O_2^{\bullet(-)}$ is that through electron transfer from the reduced form of PSs to $^3O_2$ by reaction (2):

$$PS^{\bullet(-)} + {}^{3}O_2 \rightarrow {}^{1}PS + O_2^{\bullet(-)} \qquad (2)$$

And the $PS^{\bullet(-)}$ needed in the above reaction, can be formed the self-reaction of PSs in the $T_1$ or $S_0$ state, as shown in reaction (3) and (4):

$$^{3}PS + {}^{1}PS \rightarrow PS^{\bullet(-)} + PS^{\bullet(+)} \qquad (3)$$

$$^{3}PS + {}^{3}PS \rightarrow PS^{\bullet(-)} + PS^{\bullet(+)} \qquad (4)$$

To make reaction (3) and (4) possible, the following conditions should be satisfied:

$$\mathrm{VEA}(T_1) + \mathrm{VIP}(S_0) < 0 \qquad (3\text{-a})$$

$$\mathrm{VEA}(T_1) + \mathrm{VIP}(T_1) < 0 \qquad (4\text{-a})$$

**Table 4.** VEA and VIP of $S_0$ and $T_1$ States for studied Complexes computed in DMSO using PBE0/6-31G +(d)/SDD.

| Mol. | VIP(S0) | VEA(S0) | VIP(T1) | VEA(T1) |
|---|---|---|---|---|
| **a1** | 5.73 | -2.65 | 5.76 | -3.10 |
| **a2** | 5.72 | -2.66 | 5.76 | -3.10 |
| **a3** | 5.74 | -2.56 | 5.73 | -2.98 |
| **a4** | 6.11 | -2.72 | 6.11 | -3.17 |
| **a5** | 5.60 | -2.67 | 5.45 | -2.68 |
| **a6** | 5.48 | -2.74 | 5.30 | -2.73 |
| **b1** | 5.74 | -2.88 | 5.74 | -3.15 |
| **b2** | 5.75 | -3.06 | 5.76 | -3.38 |
| **b1-r** | 5.74 | -2.60 | 3.86 | -4.91 |
| **b2-r** | 5.74 | -2.55 | 3.89 | -4.82 |
| **$^3O_2$** | | -3.40 | | |

It can be seen from **Table 4** that only **b1-r** and **b2-r** meet the energetic conditions for electron transfer (reaction 4), suggesting that these two complexes can simultaneously engage in Type I and Type II pathways. In contrast, the corresponding regioisomers **b1** and **b2** only satisfy the Type II requirement, proceeding mainly via energy transfer pathway.

Moreover, the electron and hole distribution diagrams for the $T_1$ states (Table S10) reveal that the dual photosensitization behavior of **b1-r** and **b2-r** arises from their structural features. For b1-r and b2-r, the adjustment of the phenyl ring connection position on the N^N ligand extends the hole-electron distribution to the substituents and makes the distribution more uniform, thereby increasing molecular delocalization. This facilitates electron transfer (Type I) while maintaining sufficient triplet energy for singlet oxygen generation (Type II). In contrast, the regioisomers **b1** and **b2**, lacking this connectivity, only exhibit Type II reactivity. These findings demonstrate that subtle adjustments in ligand connectivity and π-conjugation can modulate the balance between Type I and Type II pathways, providing a general strategy for designing dual-function photosensitizers.

### 3.6. Solubility, Lipophilicity, Cellular Uptake and Mitochondrial Localization

The effectiveness of PDT is influenced by cell uptake and the targeting ability in specific organelles of the PS [68–70].

Several factors, such as lipophilicity, molecular size, water solubility, and uptake mechanisms, have been reported to play important roles in the cellular uptake of Ir (III) complexes [35]. Experimental studies have shown that photosensitizer molecules with better solubility and lipophilicity exhibit significantly higher cellular uptake efficiency [55,71,72]. Due to the unique structural features of mitochondria, such as the negative membrane potential and lipid bilayer, more lipophilic cationic photosensitizers demonstrate enhanced mitochondrial targeting capacity [47,73–75].

**Table 5.** The calculated ΔGsolv (Kcal/mol) in DMSO and water solvent for all complexes at M052X/6-31G(d)/SDD level with SMD model.

| Mol. | a1 | a2 | a3 | a4 | a5 | a6 | b1 | b2 | b1-r | b2-r |
|---|---|---|---|---|---|---|---|---|---|---|
| **$\Delta G_{solv}$ (water)** | -49.79 | -48.46 | -48.95 | -47.68 | -48.61 | **-52.05** | **-50.43** | **-50.95** | **-50.47** | **-51.27** |
| **ΔGsolv (DMSO)** | -62.71 | **-63.12** | **-63.06** | -60.73 | -60.97 | **-65.75** | **-64.33** | **-65.83** | **-64.31** | **-66.08** |

To assess solvent stabilization, we calculated their solvation-free energies (ΔGsolv) in both water and DMSO using the SMD model, with the specific values provided in **Table 5**. Although solvation free energy is not directly equivalent to experimental solubility, more negative ΔGsolv values indicate thermodynamically more favorable solvent stabilization within the same computational framework. However, this relationship is not simply linear; the actual solubility is also influenced by various factors, including intermolecular interactions, temperature, and the properties of the solvent. In general, improved solvent stabilization may lead to better solubility, which is beneficial for the transport and distribution of photosensitizer (PS) molecules in biological systems and may consequently improve their bioavailability [17,55]. In our study, compared with **a1**, complexes **a2**, **a3**, **a6**, **b1**, **b2**, **b1-r**, and **b2-r** exhibit comparable or slightly more negative ΔGsolv values, indicating similar or somewhat improved solvent stabilization. These changes can be attributed to the incorporation of additional aromatic units into the **C^N** and **N^N** ligands, which extend the π-conjugation of the complexes and increase their molecular polarizability, thereby enhancing solvent stabilization.

Besides, to estimate the lipophilicity of the studied complexes, the octanol-water partition coefficient LogP of the complexes was calculated. The partition coefficient LogP of organic compounds between lipid and water phases typically refers to their distribution coefficient between n-octanol and aqueous phase, with its magnitude expressed as a logarithm[76,77], denoted as LogP. $LogP = \log C_{n\text{-}octanol} / C_{water}$. Here, $C_{n\text{-}octanol}$ and $C_{water}$ are the concentrations of compounds at equilibrium between n-octanol and water phases, respectively. To verify the feasibility of used calculation method for LogP, we selected structurally similar molecules with known experimental LogP values from the same research group [34,78]. Unlike the experiments, we only considered the effect of cationic complexes on lipophilicity, ignoring the influence of counter anions (hexafluorophosphate). To enable comparison with experimental measurements and account for counterion effects, we compared the lipophilicity of a series of experimentally synthesized Ir(III) photosensitizer with theoretically calculated lipophilicity values for their cations, subsequently deriving a fitting formula. The specific structure and fitting curve are shown in **Figure S3**~**S**4 ($y = 0.2089x + 0.5241$, $R^2 = 0.967$), and the experimental values and theoretical values of their LogP are listed in **Table S12**. In addition to the correlation coefficient, the regression model exhibits a mean absolute error (MAE) of 0.05 log units and a root mean square error (RMSE) of 0.06 log units between predicted and experimental values, indicating good agreement within the calibration set. Nevertheless, given the limited size of the calibration dataset (six complexes) and the omission of counterion effects, small predicted differences should be interpreted with appropriate caution. Therefore, we calculated the lipophilicity of designed complexes in the same way, and then estimated their lipophilicity by linear equation, the specific values are listed in **Table 6** and **Figure 8**.

**Table 6.** Calculated Lipophilicity of Studied Complexes by M052X/6-31G(d)/SDD.

| Mol | a1 | a2 | a3 | a4 | a5 | a6 | b1 | b2 | b1-r | b2-r |
|---|---|---|---|---|---|---|---|---|---|---|
| $E_{otc}$ - $E_{wat}$ (J/mol) | -31511 | -40850 | -40162 | -34607 | -30311 | -35092 | -35657 | -39797 | -35016 | -39086 |
| **LogP** (DFT) | 6.03 | 7.81 | 7.68 | 6.62 | 5.80 | 6.71 | 6.82 | 7.61 | 6.70 | 7.47 |
| **LogP** (estimated) | 1.78 | 2.16 | 2.13 | 1.91 | 1.73 | 1.93 | 1.95 | 2.11 | 1.92 | 2.09 |

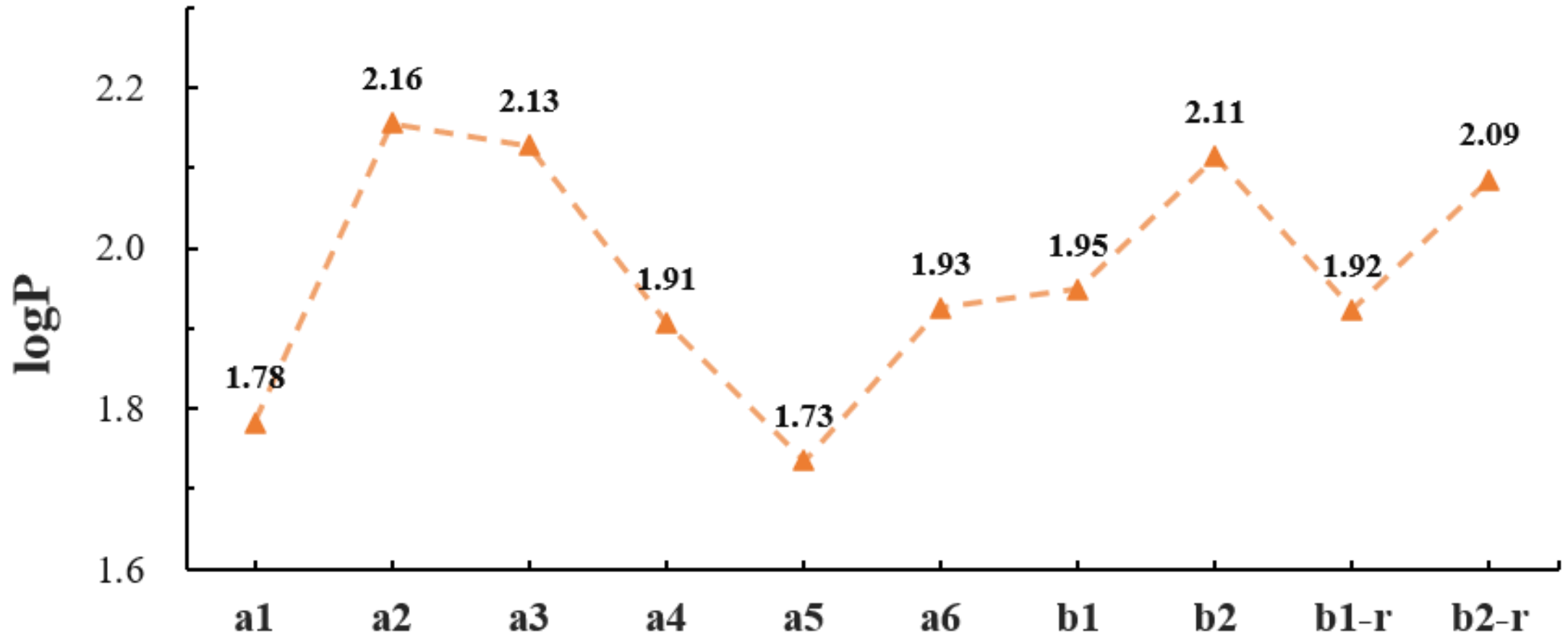

**Figure 8.** Estimated logP values of Studied Complexes.

Studies have reported that moderately lipophilic compounds (LogP 0~3) have a good balance between solubility and permeability, which is usually beneficial to enter cells [79]. Designed complexes with extended π conjugation, **a2, a3, a4, a6, b1, b2, b1-r and b2-r** (LogP ： 1.91 - 2.16), show moderately increased predicted LogP values than the experimental complex **a1** (LogP = 1.7578). In line with previous studies, LogP increases with the enlargement of π conjugation in the ligands [32]. Conversely, **a5** (LogP = 1.73) with thiophene-substituted **C^N** ligands exhibit reduced LogP value compared to **a1**. Although the variations in logP among the designed complexes are moderate, all values fall within a range commonly associated with balanced solubility and membrane permeability. The present computational analysis therefore supports relative trends in lipophilicity within a consistent theoretical framework.

Based on the combined solubility and lipophilicity analysis, complexes **a2, a3, a6, b1, b2, b1-r and b2-r** exhibit physicochemical characteristics that may favor membrane interaction compared to **a1**. Previous literature reported that cationic molecules with logP values between 0 and +5 tend to show mitochondrial localization. [47,80]. It has been reported that complex **a1** targets mitochondria with a Pearson's correlation coefficient of r = 0.547. Within the context of these empirical observations, the moderately increased predicted lipophilicity of the designed complexes suggests a potential for enhanced mitochondrial affinity. However, such biological implications remain hypothetical and require experimental validation.

## 4. Conclusion

In this study, we designed a series of iridium-based two-photon photosensitizers by modifying the chelated and auxiliary ligands, thereby improving key photophysical properties such as two-photon absorption (TPA), triplet-state lifetimes, and lipophilicity. The fused-π extension modification to the chelated ligands and ancillary ligands reduce the electron-accepting or -donating ability of both ligands in two-photon transition, causing the two-photon transition type to change from MLCT to ILCT and resulting in the enhanced two-photon absorption in **a2**, **b2** and **b1-r**. Especially in TP - PDT efficiency (such as two-photon absorption (TPA), triplet-state lifetimes, and lipophilicity), this indicates that the asymmetric iso-fused benzene ring modification of the **N^N** ligand is a robust design strategy for comprehensively enhancing photosensitization.

1. The TPA cross-sections of **a2**, **b2**, **b1-r**, and **b2-r** were significantly enhanced relative to **a1** (13.72 GM), with values ranging from 15.53 GM to 56.51 GM. These improvements demonstrate the successful optimization of the TPA properties through the fused-benzene modification.
2. The designed complexes, except for **b2-r**, exhibited extended triplet-state lifetimes (ranging from2.62 to 85.96 μs), longer than that of **a1** (2.19 μs). These extended triplet lifetimes will helpful to enhance the efficiency of singlet oxygen generation, which is crucial for PDT efficacy.
3. The fused-benzene ring on the **N^N** ligand improves solubility and lipophilicity. Compounds **a2**, **a6**, **b1**, **b2**, **b1-r** and **b2-r** show better water solubility and enhanced lipophilicity than **a1**, which may promote cellular uptake and improve mitochondrial targeting.
4. Among all molecules, **b1-r** and **b2-r** have the lowest VIP and the largest VEA, making them making them more likely to act as electron donors and least likely to act as electron acceptors. This implies that, with respect to $O_2$, these complexes with the larger VEA may facilitate electron-transfer reactions with it, which could potentially enable Type I ROS generation pathways. Such behavior could be advantageous for addressing hypoxic tumor environments, where Type II mechanisms can be less efficient.

In summary, compared to the reference molecule a1, the complexes **a2**, **b2**, and **b1-r** showed improvements in TPA cross-sections, triplet-state lifetimes, and lipophilicity, making them promising PS candidates for two-photon PDT. It is worth noting that, within the theoretical framework, the **b1-r** complex is capable of undergoing PDT photosensitization via both type I and type II reaction mechanisms. This is expected to be further verified experimentally. We hope this work can provide valuable theoretical guidance for the study of two-photon absorption photosensitizers based on metal Ir complexes.

**ACKNOWLEDGMENTS**

This work is supported by Science and Technology Development Plan Project of Jilin Province，China（Nos.20240101167JC), the Natural Science Foundation of China (nos. 21473071, 21173099, 20973078 and 20673045).

**Graphic for manuscript**

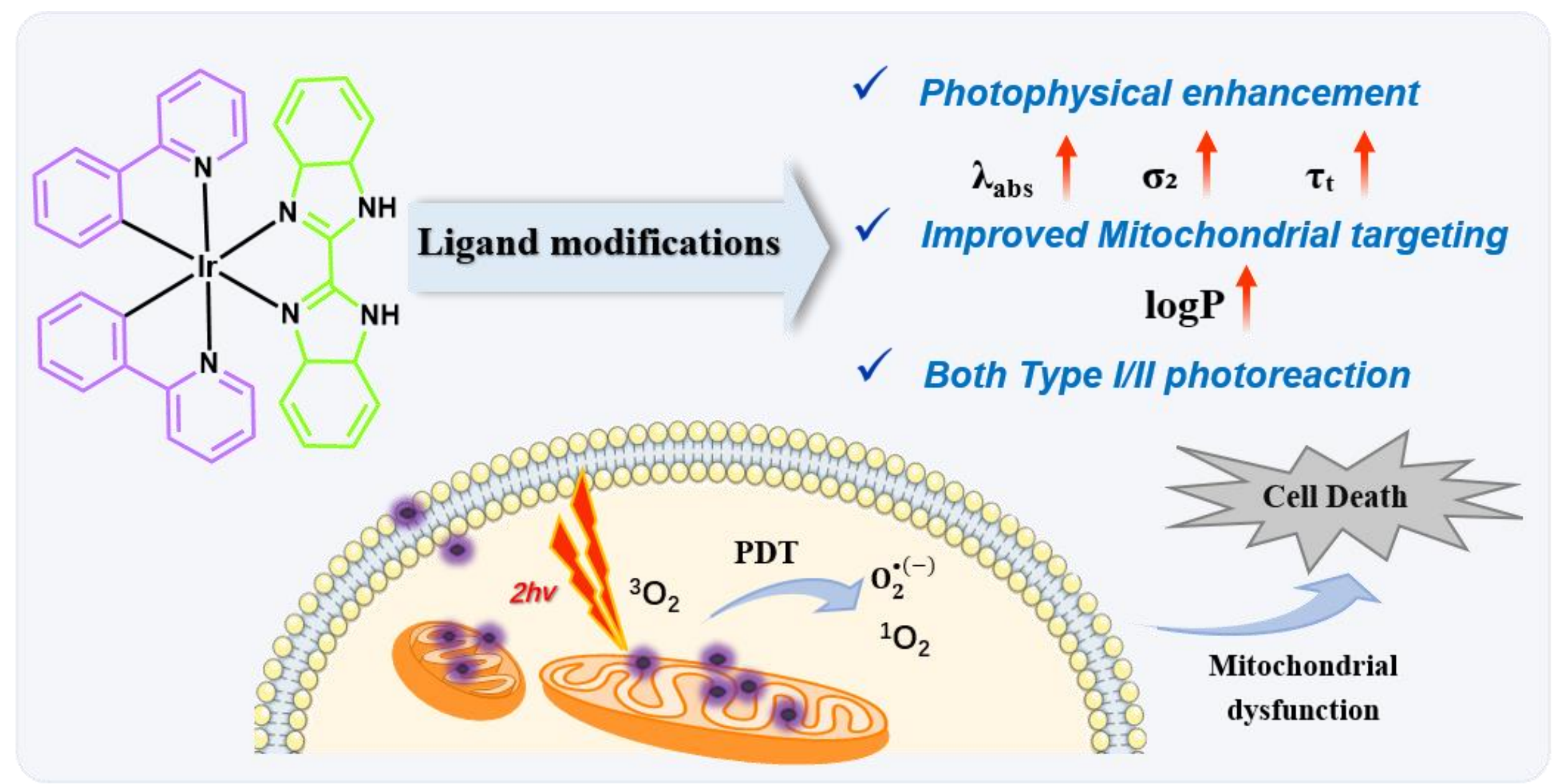